\documentclass[a4paper]{article} 
\pdfoutput=1
\usepackage{geometry}
\usepackage{setspace}
\usepackage{multicol}
\usepackage[square, numbers, comma, sort&compress, sectionbib]{natbib} 
\usepackage{textcomp}
\usepackage{graphicx}
\usepackage{xcolor}
\usepackage{epstopdf}
\AppendGraphicsExtensions{.tiff}
\usepackage{everyshi}
\usepackage{eso-pic}
\usepackage{ifthen}
\usepackage{settobox}
\usepackage{lscape}
\usepackage{supertabular}
\usepackage[stable,symbol]{footmisc}
\usepackage[utf8]{inputenc}
\usepackage[english]{babel}
\usepackage{amsmath}
\usepackage{amsfonts}
\usepackage{amssymb}
\usepackage{graphicx}
\usepackage{epstopdf}
\usepackage{float}
\usepackage{color}
\usepackage[font=small,labelfont=bf]{caption}
\newcommand{\sn}[1]{{\color{black}{#1}}}
\usepackage{authblk}

\def\be{\begin{equation}}
\def\ee{\end{equation}}
\def\bea{\begin{eqnarray}}
\def\eea{\end{eqnarray}}
\def\ba{\begin{array}}
\def\ea{\end{array}}

\begin{document}

\title{Porosity governs normal stresses in polymer gels}

\author[a]{Henri C. G. de Cagny\footnote{authors contributed equally}}
\author[b]{Bart E. Vos$^*$}
\author[c]{Mahsa Vahabi$^*$}
\author[b,d]{Nicholas A. Kurniawan}
\author[e]{Masao Doi}
\author[b]{Gijsje H. Koenderink \footnote{Email: g.koenderink@amolf.nl}}
\author[c]{Fred C. MacKintosh \footnote{Email: fcmack@gmail.com}}
\author[a]{Daniel Bonn \footnote{Email: d.bonn@uva.nl}}

\affil[a]{Institute of Physics, University of Amsterdam, Amsterdam, The Netherlands}
\affil[b]{FOM-Institute AMOLF, Amsterdam, The Netherlands}
\affil[c]{Department of Physics and Astronomy, VU Universiteit Amsterdam, The Netherlands}
\affil[d]{Department of Biomedical Engineering \& Institute for Complex Molecular Systems, Eindhoven University of Technology, Eindhoven, The Netherlands}
\affil[e]{Center of Soft Matter Physics and its Applications, Beihang University, Beijing, China }

\date{}

\maketitle

\textbf{When sheared, most elastic solids such as metals, rubbers and polymer hydrogels dilate in the direction perpendicular to the shear plane. This well-known behaviour known as the Poynting effect is characterized by a positive normal stress \cite{Poynting1909}. Surprisingly, biopolymer gels made of fibrous proteins such as fibrin and collagen, as well as many biological tissues exhibit the opposite effect, contracting under shear and displaying a \emph{negative} normal stress \cite{Janmey2007,horgan2015}. Here we show that this anomalous behaviour originates from the open network structure of biopolymer gels, which facilitates interstitial fluid flow during shear. Using fibrin networks with a controllable pore size as a model system, we show that the normal stress response to an applied shear is positive at short times, but decreases to negative values with a characteristic time scale set by pore size. Using a two-fluid model, we develop a quantitative theory that unifies the opposite behaviours encountered in synthetic and biopolymer gels. Synthetic polymer gels are impermeable to solvent flow and thus effectively incompressible at typical experimental time scales, whereas biopolymer gels are effectively compressible. Our findings suggest a new route to tailor elastic instabilities such as the die-swell effect that often hamper processing of polymer materials and furthermore show that poroelastic effects play a much more important role in the mechanical properties of cells and tissues than previously anticipated.}

\vspace{5mm}

When subjected to a shear stress, materials either shrink (shear contraction) or expand (shear dilatancy). As shown by Poynting more than a century ago \cite{Poynting1909}, simple elastic solids exhibit shear dilatancy. Similar behaviour has since been observed in more complex viscoelastic systems, such as granular materials, rubbers and polymer glasses \cite{Reynolds1885,Larson1998}. \sn{The tendency of such materials to dilate can be measured as a positive normal compressive stress or pressure that develops when a sample is sheared between two plates with a fixed gap.} In case of polymer materials, shear dilatancy is usually described by the classical Mooney-Rivlin model \cite{Mooney1940,Rivlin1948}, which predicts a normal stress difference $N_1 \sim G\gamma^2$, where $\gamma$ is the shear strain and $G$ the network shear modulus. In Fig.~1a this behaviour is illustrated for polyacrylamide {(PAAm)} hydrogels of varying stiffness subjected to an oscillatory shear deformation. Surprisingly, biopolymer networks have been reported to exhibit the opposite response, contracting when sheared \cite{Janmey2007, Kang2009,horgan2015}. This behaviour is clearly illustrated in Fig.~1b, which shows that aqueous gels of the blood clotting protein fibrin develop a negative normal stress under shear. The magnitude of the normal stress again increases quadratically with strain, but it reaches values comparable to the shear modulus at significantly lower shear strain ($\gamma\simeq 1/10$) than for polyacrylamide ($\gamma\simeq1$). The origin for the remarkable difference in the sign and magnitude of the normal stress between synthetic hydrogels and biopolymer gels is still unknown.

\begin{figure}[htbp]
\centering
\includegraphics[width=0.5\textwidth]{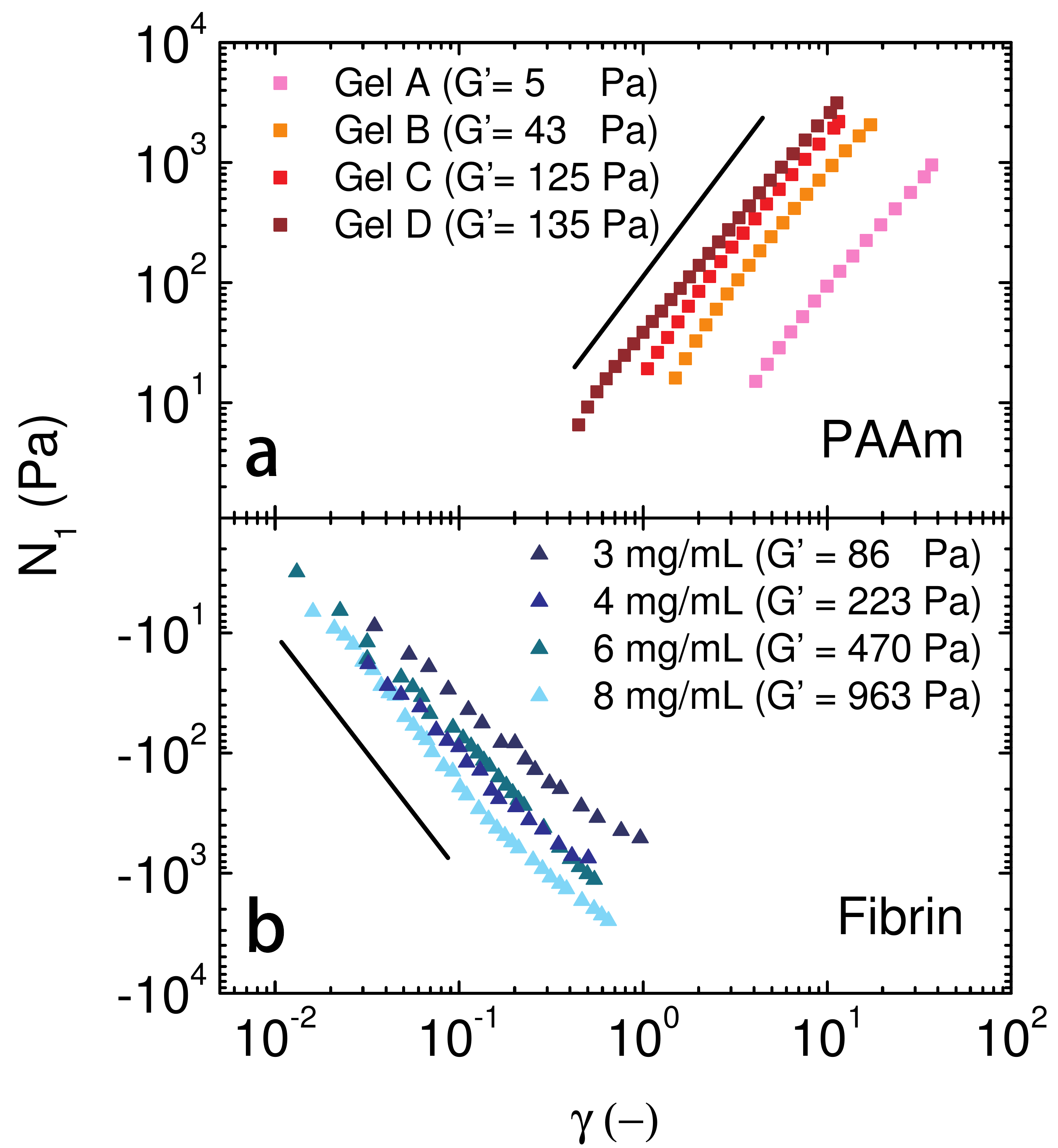}
\caption{Normal stress difference $N_1=\frac{2F}{\pi R^2}$, where $F$ is the normal force (thrust) reported by the rheometer and $R$ is the sample radius, as a function of the amplitude of the applied oscillatory shear strain for \textbf{(a)} PAAm \cite{SI} prepared with various ratios of monomer-to-cross-linker concentrations. The line indicates a quadratic dependence of $N_1\sim\gamma^2$, as expected from the Mooney-Rivlin model \cite{Mooney1940,Rivlin1948}. In \textbf{(b)}, {$\frac{2F}{\pi R^2}$} is shown for fibrin gels polymerized at 22\textdegree C at various fibrinogen concentrations (in mg/mL). The line indicates a {$\sim\gamma^2$ dependence,} but with negative sign.}
\end{figure}

{Here, we aim to understand the mechanistic basis of the fundamentally different response of synthetic and biopolymer gels and to develop a minimal model that can capture the behaviour of both types of gels.
In either case, the normal stress response is fundamentally nonlinear, since its sign cannot reverse when the shear train $\gamma$ is reversed. Thus, to lowest order, normal stress is expected to vary as $\gamma^2$, even while the shear stress remains linear in $\gamma$. Although both gels in Fig.\ 1 are hydrogels containing over $90\%$ interstitial water, structurally there is a profound difference in the pore size. While polyacrylamide gels have a pore size of order 10 nanometers \cite{tombs1965}, fibrin networks have pore sizes that can be in the micrometer range \cite{Okada1985,Pieters2012,Lang2013}. Fluid permeability can therefore play an important role in the mechanical response.} For hydrogels with a small pore size, we expect a strong viscous coupling between the network and the solvent, which will suppress motion of the network relative to the solvent and effectively render the gel as a whole incompressible. By contrast, biopolymer gels can expel interstitial fluid to relax pressure gradients on long enough time scales, allowing the network to contract upon shearing \cite{heussinger2007nonaffine,Conti2009,Kang2009,Tighe2013}.

\begin{figure}[htbp]
\centering
\includegraphics[width=0.7\textwidth]{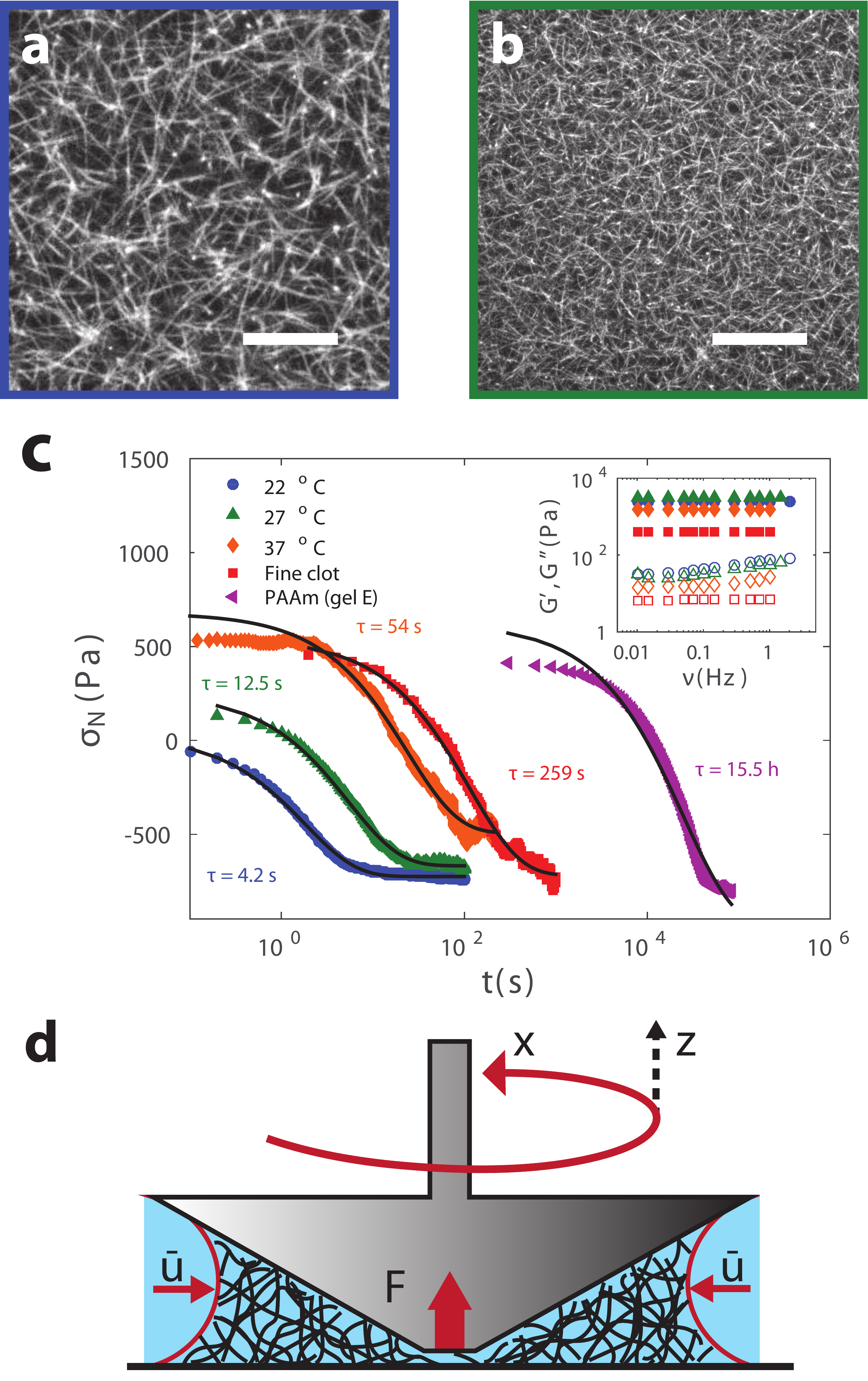}
\caption{Fluorescence confocal microscopy images of fibrin networks whose pore size is tuned by polymerizing under different conditions: at 22\textdegree C \textbf{(a)} and 27\textdegree C \textbf{(b)}. The scale bars are $10~\mu$m. Protein content is (8 mg/mL) in both samples.
\textbf{(c)}Normal stress $\sigma_N$, given by the apparent normal stress difference $\frac{2F}{\pi R^2}$ obtained from the rheometer thrust $F$, for four fibrin networks differing in pore size as a function of time after the application of a constant shear stress at $t=0$. The stress relaxation curves are fitted to an exponential decay derived from the two-fluid model in \cite{SI} (black lines). \sn{The viscoelastic time scale is unrelated to the normal stress transition, as shown in the inset where the storage moduli (filled symbols) and the loss moduli (open symbols) of the gels are plotted.} \textbf{(d)} Schematic representation of the two-fluid model showing an inward, radial contraction of the network (black) relative to the solvent (blue) upon shearing.}
\end{figure}

{To study the role of porosity, we choose fibrin networks as a model system, the pore size of which can be tuned} from nanometers to micro\-meters by simply changing the temperature, ionic strength and pH during self-assembly \cite{Nunes1995}. This is demonstrated in Figs.~2a and b, which show fluorescence microscopy images of two fibrin gels that are assembled at the same monomer concentration of 8 mg/mL but at different temperatures. Using quantitative measurements of the fiber mass-length ratio by light scattering, we calculate average mesh sizes of $0.36\ \mu$m and $0.29\ \mu$m for these networks (see \cite{SI} for further details).

To test the influence of pore size on the normal stress response, we subject each network to a constant shear stress and we monitor the normal stress as a function of time. Intriguingly, we find that in each case, the normal stress relaxes from an initially positive or close-to-zero value to a negative steady-state value with a rate that strongly varies with pore size. The characteristic relaxation time, $\tau$, increases from just a few seconds to $\sim$100~s as the pore size {of fibrin} decreases from $0.36\ \mu$m to $0.08\ \mu$m. {For PAAm gels with a pore size $\sim10$~nm \cite{tombs1965}, the relaxation time grows to over 15 hours (Fig.~2c).} These observations support our hypothesis that the sign of the normal stress is controlled by the time scale for solvent flow through the network. The data suggest that the normal stress is positive as long as the polymer network and the fluid remain viscously coupled, and switches sign to become negative when the fluid can move relative to the network. Importantly, the timescale separating these behaviours is unrelated to the timescales apparent in the linear viscoelastic response (see Fig. 2c).

To quantitatively model the effects of network poroelasticity in the shear rheology of polymer networks, we start from the two-fluid model \cite{Brochard1977,Milner1993,Gittes1997,Levine2000} that describes a polymer gel as a biphasic system comprised of a linear elastic network immersed in a viscous and incompressible liquid \cite{SI}. The two components are coupled by a force per unit volume, $\Gamma\left(\dot{\vec{u}}-\vec v\right)$, acting on the liquid and opposite to the force on the network. This dissipative force arises from the relative motion of the solvent, which moves with velocity $\vec v$ and the network, with velocity $\dot{\vec{u}}$. For a network with pore size $\xi$ and a fluid with viscosity $\eta$, $\Gamma \sim \eta/\xi^2$, {since the Stokes drag force on a network strand of size $\sim\xi$ moving with relative velocity $\sim v$ is $\sim\eta\xi\Delta v$ and this acts on a volume $\sim\xi^3$ \cite{SI}. Given the small polymer volume fraction $\phi$ of most hydrogels and biopolymer networks ($\phi\sim10^{-3}$ for fibrin gels), the radial velocity component of the incompressible fluid effectively vanishes and the only radial motion is due to the network.} This radial motion, $\dot u_r$, generates a radial pressure gradient in the solvent given by:
\be
\nabla_rP=\Gamma\dot u_r=-\tilde\sigma/r-(K/r^2)u_r,\label{2Fluidshort}
\ee
for a cone-plate geometry.
The net force on the network has two distinct elastic contributions. The first contribution comes from the hoop stress $\tilde{\sigma}$, which tends to drive the network radially inward (Fig.\ 2d) \cite{SI}: hoop stresses generated by shearing tend to drive radial contraction of the network and expulsion of the solvent, {much as a twisted sponge expels water by contracting radially.}
{By symmetry, $\tilde\sigma\sim\gamma^2$ to lowest order, as noted above, although Eq.\ \eqref{2Fluidshort} is linear in $u_r$.}
The second contribution to the net force on the network comes from a restoring force that balances the radial contraction on long time scales (i.e., as $\dot u_r\rightarrow0$). This restoring force originates from the gradient in the elastic shear stress $\sim G\nabla_z u_r$ that results from the axial ($z$) variation of $u$ (see Fig.\ 2d).
For a cone-plate rheometer with small gap size $d$ and small cone angle $\alpha$, the restoring force $\sim G/d^2u_r $. Thus, since $d=\tan(\alpha) r$, $K\sim G/\tan(\alpha)^2$ in Eq.\ \eqref{2Fluidshort}. We thus predict a characteristic relaxation time $\tau\sim \eta d^2/G\xi^2$. Indeed, we experimentally observe a rapid decrease of the relaxation time with increasing pore size, consistent with the predicted scaling (see Fig. S1 in \cite{SI}).

We can consider two opposite limits of Eq. \eqref{2Fluidshort}.
{In the limit of small pore size and $\Gamma\rightarrow\infty$, the radial displacement $u_r\rightarrow0$ (with finite $\Gamma\dot u_r$) and Eq.\ \eqref{2Fluidshort} reduces to $\nabla_rP=-\tilde\sigma/r$.} Shearing will thus increase the pressure toward the axis of the rheometer, which results in a positive contribution to the normal force. Dense hydrogels will therefore effectively behave as incompressible materials for which the normal force $F$ is related to the normal stress difference $\sigma_{xx}-\sigma_{zz}$ by $N_1=2F/\pi R^2$, where $R$ is the sample radius \cite{Venerus2007}. $N_1$ is positive for a rubber-like material, consistent with measurements on polyacrylamide gels \cite{Rivlin1948}. In the opposite limit of networks with a large pore size, the pressure difference can relax by water efflux and in steady state the two terms on the right hand side of Eq.\ \eqref{2Fluidshort} cancel. \sn{In the absence of this pressure, only the polymer stress terms remain and the normal force measured by the rheometer is given by $F=-\pi R^2\sigma_{zz}$, corresponding to a reported (apparent) normal-stress difference $N_1^{\mbox{\rm\scriptsize (app)}}=-2\sigma_{zz}$.}

A key prediction of the two-fluid model is that the \sn{response} of the normal stress measured in a rheology experiment should depend on the experimental time scale relative to the characteristic relaxation time, $\tau\sim \eta d^2/G\xi^2$. To quantitatively test this prediction, we subject the fibrin gels to an oscillatory shear stress with frequencies between 0.001 Hz and 5 Hz, allowing us to conveniently probe a range of time scales from 0.2 to 1000 s in a single experiment. We measure the normal stress response after the system has reached steady state (Fig. S2 in \cite{SI}). \sn{We focus on fibrin gels polymerized at 27\textdegree C, which have a relaxation time $\tau\approx12.5$ s that lies in the middle of the experimentally accessible frequency range. In steady state, the observed normal stresses are indeed negative over the entire frequency range.}

When we plot the time-dependent normal stress (Fig.\ 3d-f) as a function of shear stress, we obtain the Lissajous curves shown in Fig.\ 3a-c. Strikingly, the Lissajous curves completely change with changing frequency. For oscillation periods longer than $\tau$ (Fig. 3a), the normal stress decreases with increasing shear stress, demonstrating contractile behaviour under shear. By contrast, for oscillation periods shorter than $\tau$, the normal stress increases with increasing shear stress, demonstrating extensile behaviour (Fig. 3c). The transition occurs at an intermediate frequency that is of order $1/\tau$ (Fig. 3b). This experiment unambiguously shows that the normal stress response of a polymer gel is governed by fluid flow, which is suppressed at higher frequencies. The normal stress response is therefore controlled by the network pore size, and is furthermore dependent on the shear modulus $G$ and the gap size $d$ between the cone and the plate (see Fig. S3 in \cite{SI}).

The two-fluid model allows us to perform an even more rigorous test of the mechanism governing the normal stress response of polymer gels, since we can calculate the time-dependence of the normal stress and compare it to experiments. For symmetry reasons, the normal-stress components $\sigma_{xx}$ and $\sigma_{zz}$ are expected to have a leading $\gamma^2$ dependence on strain. Since the shear stress $\sigma_{xz}\simeq G\gamma$, we define
$\sigma_{xx}\equiv A_xG\gamma^2\quad\mbox{and}\quad\sigma_{zz}\equiv A_zG\gamma^2$. For an oscillatory strain $\gamma(t)=\gamma_0\sin(\omega t)$, the steady state solution of the time-dependent Eq.~\eqref{2Fluidshort} is \cite{SI}
\begin{equation}
N_1^{\mbox{\scriptsize (app)}} =-2A_zG\gamma(t)^2+\tilde AG\gamma_0^2({\mathcal A}\cos (2 \omega t) +{\mathcal B}\sin (2\omega t)),\label{Napp}
\end{equation}
where
\begin{equation}
{\mathcal A}=-\frac{1}{8 \omega\tau}
\Big[2 \tan ^{-1}\left(1+2 \sqrt{\omega\tau}\right) +2 \tan ^{-1}\left(1-2 \sqrt{\omega\tau}\right)-\pi+4 \omega\tau\Big],\label{calA}
\end{equation}
\be
{\mathcal B}=\frac{1}{8\omega\tau }\log\left(1+4\omega^2\tau^2\right),\label{calB}
\ee
with $\tilde A=A_x-A_z$ and $\tau=\Gamma R^2/K \sim\eta d^2/G\xi^2$. In the limit where $\omega\tau\gg1$, Eq.~\eqref{Napp} reduces to the well-known Mooney-Rivlin expression for incompressible materials, $N_1=G\gamma^2$. In the opposite limit where $\omega\tau\ll1$, Eq.~\eqref{Napp} instead reduces to $N_1^{\mbox{\scriptsize (app)}}=-2A_zG\gamma^2$.

Based on prior measurements on a range of biopolymer gels in the $\omega\tau\ll1$ limit \cite{Janmey2007,Kang2009, Storm2005} as well as models of fibrous networks \cite{MacKintosh1995,Storm2005,heussinger2007nonaffine,Conti2009}, we anticipate $A_{z}\sim 1/\gamma_c$, where $\gamma_c$ is the onset strain for nonlinear elasticity, which is typically $\sim 1/10$. Thus, in the limit of low frequencies, not only is $N_1^{\mbox{\rm\scriptsize (app)}}$ negative, but its magnitude can actually be much larger than $\sigma_{xy}\gamma$.

\begin{figure*}[htbp]
	\centering
	\noindent\makebox[\textwidth]{\includegraphics[width=16cm]{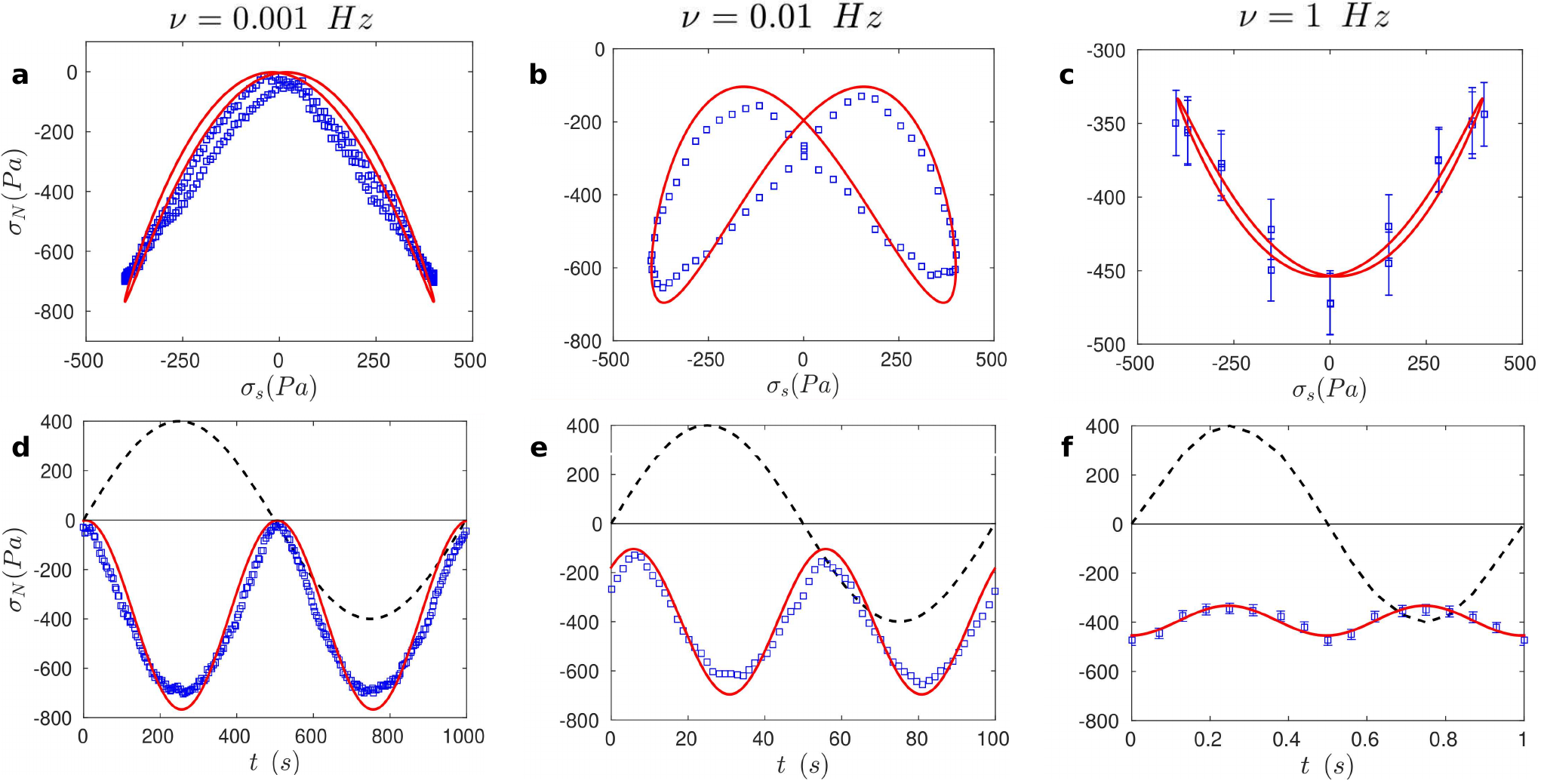}}
	\caption{Normal stress $\sigma_N$, given by the apparent normal stress difference $N_1^{\mbox{\rm\scriptsize (app)}}=\frac{2F}{\pi R^2}$ reported by the rheometer, for a fibrin gel polymerized at 27\textdegree C in response to an oscillatory shear stress at oscillation frequencies of 0.001 Hz (left), 0.01 Hz (middle) and 1 Hz (right). The top panels \textbf{(a)-(c)} show the Lissajous curves of normal stress versus shear stress (symbols) fitted by the predictions of the two-fluid model in Eq.~2 (red lines) assuming a time constant of 12.5~s. The bottom panels \textbf{(d)-(f)} show the corresponding time-dependent normal stress (blue open symbols) and applied shear stress (black dotted lines). The data shown at $\nu = 1$ Hz represent averages with standard deviations obtained by averaging over 34 cycles to compensate for the low sampling frequency of the rheometer. The normal stresses are all negative since they correspond to steady-state values, obtained after initial relaxation (Fig.\ S2 in \cite{SI}).}
\end{figure*}

To test these predictions experimentally, we fit the oscillatory normal-stress data shown in Fig.~3 to Eq. (2). The only fit parameters are $A_z$ and $\tilde{A}$, since the shear modulus $G$ is measured independently from the shear stress at small strain and the relaxation time $\tau$ is measured independently from the normal stress relaxation upon applying a constant shear stress ($\tau = 12.5$ s, Fig.~2c). We observe excellent agreement between the data (symbols) and the model (solid lines) over the entire range of oscillation frequencies (Fig.\ 3), with fitting parameters that are insensitive to frequency (Fig.\ S1 in \cite{SI}).

Our observations reveal that poroelastic effects involving interstitial fluid flow play an unexpectedly important role in the shear rheology of polymer gels. Poroelastic effects in porous media such as fluid-imbibed polymer gels are usually considered to affect only volume-changing deformations such as compression and extension \cite{Biot1956,Casares2015,Oosten2016}. Our experiments and theory demonstrate that the shear response of polymer gels is highly sensitive to fluid flow and network compressibility, in spite of the volume-conserving nature of simple shear deformations. Depending on the timescale of deformation and the hydrodynamic coupling of the polymer network with the surrounding solvent, polymer gels behave as either incompressible materials with a positive normal stress or compressible materials with a negative normal stress. 

We demonstrated that the normal stress response of both synthetic and biopolymer gels is quantitatively captured by a minimal model that takes into account the biphasic nature of hydrogels. This model can explain why synthetic hydrogels exhibit shear-dilation, while biopolymer gels have been reported to exhibit shear-contraction. This suggests a new route to tailor the sign and magnitude of the normal stresses for polymer materials by tuning the pore size, solvent viscosity, and nonlinear shear elasticity. This could prove valuable in the context of materials science, since normal stresses can cause elastic instabilities that severely complicate processing \cite{Larson1998}. Finally, our findings highlight the important role of poroelastic effects in tissue and extracellular matrix mechanics, where normal stresses can become a dominant stress component, even for small strains of order 10\% \cite{Kang2009}. Related poroelastic effects in intracellular networks have previously been shown to govern the rheology of cells. However, the much smaller cellular dimensions $d\simeq 1~\mu$m, can be expected to limit the corresponding poroelastic relaxation time to of order 1~s, even for the smaller mesh sizes of order 10~nm \cite{Moeendarbary2013}, which renders cells effectively compressible on timescales $\gtrsim$ 1~s.

\section*{Acknowledgments} This work was supported by the Foundation for Fundamental Research on Matter (FOM), which is part of the Netherlands Organisation for Scientific Research (NWO).



\pagebreak
\begin{center}
\textbf{\Huge Supplemental Materials\\
}
\end{center}
\setcounter{equation}{0}
\setcounter{figure}{0}
\setcounter{table}{0}
\setcounter{page}{1}
\makeatletter
\renewcommand{\theequation}{S\arabic{equation}}
\renewcommand{\thefigure}{S\arabic{figure}}


\vspace{10mm}

\section*{Materials and Methods}

Chemicals were purchased from Sigma Aldrich (Zwijndrecht, the Netherlands). Human plasma fibrinogen and thrombin were purchased from Enzyme Research Laboratories (Swansea, United Kingdom). Fibrinogen stock solution was diluted to 8 mg/mL in assembly buffer (150 mM $\textrm{NaCl}$, 20 mM HEPES and 5 mM $\textrm{CaCl}_2$) at a pH of 7.4. Dense networks (fine clots) with an average pore size of 0.08 $\mu$m were obtained in fine-clot assembly buffer (400 mM $\textrm{NaCl}$ and 50 mM Tris-HCl) at a pH of 8.5 \cite{Bale1985}. The mixtures were prewarmed to the desired polymerization temperature and 0.5 U/mL thrombin was added to initiate network formation. The fibrin gels were allowed to polymerize \textit{in situ} for at least 12 hours (22\textdegree C and 27\textdegree C samples) or 4 hours (37\textdegree C samples) between the prewarmed cone and plate geometry of an Anton Paar rheometer (Physics MCR 302, Graz). Polyacrylamide gels were polymerized by preparing a mixture of polyacrylamide and N,N'-methylenebis(acrylamide)(Bis) followed by dilution to the desired final concentration. Polymerization was initiated by adding ammonium persulfate (0.5 $\mu$g/mL) and tetramethylethylenediamine (1 $\mu$L/mL) and allowed to proceed in situ between the rheometer plates at 20 \textdegree C. \sn{We used a 40 mm, 2\textdegree~stainless steel cone-plate geometry for all fibrin data and a 50 mm, 2\textdegree~stainless steel cone-plate geometry for polyachrylamid data reported in the main text}, and a series of stainless steel cone-plate and plate-plate geometries in additional experiments reported in the Supplementary Information. Solvent evaporation was prevented by adding a layer of mineral oil (Sigma Aldrich, M3516) to cover the liquid-air interface. We checked that this procedure did not influence the normal stress response. The composition of the polyacrylamide gels displayed in Figure~1a of the main text are the following:

\begin{table}[h]

\begin{center}

\begin{tabular}{|c|c|c|}
  \hline
  Gel & Total polymer mass fraction (\%) & Acrylamide-MBAA mass ratio \\
  \hline
  A & 4 & 149:1 \\
  \hline
  B & 5 & 199:1 \\
  \hline
  C & 4 & 99:1 \\
  \hline
  D & 6 & 299:1 \\
    \hline
  E & 3 & 39:1 \\
  \hline
\end{tabular}
\end{center}
\end{table}

We measured the linear elastic shear modulus $G$ of the gels by measuring the stress response to a small oscillatory shear strain with an amplitude of $0.1\%$ and frequency of 1 Hz. The normal stress response to an applied shear was obtained by applying Large-Amplitude Oscillatory Shear (LAOS) at a range of frequencies ($0.001$ Hz to $7$ Hz) at a shear stress amplitude of 800 Pa. We measured the time-resolved strain and normal stress response using an oscilloscope coupled to the analogue outputs of the rheometer. The characteristic normal-stress relaxation time was obtained by applying a constant shear stress (400 Pa for the 22 \textdegree C and the 27 \textdegree C gels, 550 Pa for the 37 \textdegree C gel, 500 Pa for the fine clot gel and a strain of 150\% for PAAm). \sn{The relaxation data for PAAm were smoothed using Savitzky-Golay filter. The relaxation stress experiment was initiated one day after the polymerization to ensure the stability of the base line}   

To characterize the pore size of the fibrin gels, we performed light scattering measurements using a spectrophotometer (Lambda35 UV/VIS Perkin Elmer, Waltham, MA, USA). Gels were polymerized in quartz cuvettes and absorbance spectra were taken over a wavelength range from 450 to 900 nm. The mass-length ratio of the fibers was obtained by fitting the spectra to a scattering model that assumes a random network of rigid cylindrical fibers \cite{Yeromonahos2010,Piechocka2016}. The average mesh size $\xi\sim\sqrt{1/\rho}$ follows from the fiber length density $\rho = \frac{\mu}{c_p}$, where $\mu$ is the fiber mass-length ratio and $c_p$ the fibrinogen concentration. To validate the light scattering measurements, we also performed image analysis of confocal fluorescence microscopy images of fibrin networks doped with 5 mole\% AlexaFluor 488-conjugated fibrinogen (Life Technologies, Eugene, Oregon, USA). Images were obtained on a Nikon Eclipse TI confocal system with a 100x oil immersion objective (NA=1.40). Image analysis is discussed in the Supplementary Information section 'Pore size analysis'.

\section*{Two-fluid model and relaxation of hoop stress}
\sn{When an elastic body is sheared by twisting, either in the classic wire experiments of Poynting or in a conventional  cone-plate or parallel plate rheometer, sections of the material tend to be stretched out circumferentially. The resulting sections also tend to develop tensile stresses as they are stretched out. Through the combination of tension and the curved shape, these tensile hoop stresses result in a net inward radial force, much as a curved rubber band would when stretched into a circular shape.  These effects contribute to a net stress $\tilde\sigma$ and radial force, proportional to the curvature of the streamlines, tending to drive inward contraction of the gel. By symmetry, this force must be even in the applied strain, since the tension is independent of the direction of rotation of the rheometer.} Thus, to lowest order, a quadratic dependence on shear strain $\gamma$ is expected. We define this force (per unit volume) to be
\be
f_r=-\frac{\tilde\sigma}{r}\simeq-\frac{1}{r}\tilde AG\gamma^2,\label{hoop}
\ee
where the coefficient $\tilde A$ is dimensionless.  
In an incompressible medium, in which net radial motion is not possible, this radial force must be balanced by a pressure that builds toward the axis of the rheometer. In the presence of a free surface, as opposed to a rheometer plate, this gives rise to rod-climbing behavior. In the case of a rheometer, the pressure results in a positive, upward thrust $F$. 

For a polymer gel consisting of both network and solvent, radial motion of the network is possible, which can lead to a relaxation of the pressure contribution to the measured thrust. To model this relaxation, we use the two-fluid model \cite{Brochard1977,Milner1993,Gittes1997,Levine2000}, in which the network displacement $u$ and solvent velocity $v$ satisfy the coupled equations. The equation for the net force per unit volume acting on the fluid in the non-inertial limit is 
\be
0=\eta\nabla^2\vec v-\vec\nabla P-\Gamma\left(\vec v-\dot{\vec{u}}\right),\label{2fluidViscous}
\ee
while the corresponding (non-inertial) equation for the force on the network is 
\be
0=G\nabla^2\vec u+(G+\lambda)\vec\nabla\cdot(\vec\nabla\cdot\vec u)+\Gamma\left(\vec v-\dot{\vec{u}}\right),\label{2fluidElastic}
\ee
where $\eta$ is the solvent viscosity, $G$ is the shear modulus and $\lambda$ is a Lam\'e coefficient that is typically of order $G$. These equations are coupled by a term representing the force on the fluid (and opposite to the force on the network) due to the relative motion of the two components. The coupling constant $\Gamma$ is expected to be of order $\eta/\xi^2$ for a network mesh size $\xi$. This can be estimated in a free-draining approximation by considering the drag force on a network strand of length $\sim\xi$ moving with velocity $\dot{\vec u}$ with respect to a stationary solvent. Apart from a weak logarithmic correction, this drag force is approximately $4\pi\xi\eta\dot u$. This is the force per mesh volume $\sim\xi^3$, meaning that the force per unit volume is of order $\eta\dot u/\xi^2$. This is the net force per unit volume on the fluid given by the last term on the right-hand side of \eqref{2fluidViscous} with $\vec v=0$. 

If the volume fraction of the network is small, as it is for most biopolymer gels ($\sim10^{-3}$), then we can safely assume that only the network moves radially in response to strain-induced hoop stresses $\tilde\sigma$, while the solvent remains stationary since it is incompressible. In this case, the radial component of \eqref{2fluidViscous} reduces to
$\nabla_r P=\Gamma\dot{u_r}$. Again, the radial motion of the network at low volume fraction is well approximated by
\be
\Gamma{\dot u_r}\simeq G \frac{\partial^2}{\partial z^2}u_r.\label{urcomp1}
\ee
Here, we have assumed a cone-plate geometry with gap $d\ll r$. Corrections to \eqref{urcomp1}, from both $\nabla^2\vec u$ and $(G+\lambda)\vec\nabla\cdot(\vec\nabla\cdot\vec u)$ terms are smaller by of order $(d/r)^2$. Together with the boundary conditions that $u_r=0$ at both $z=0$ and $z=d$, we find a characteristic relaxation time 
\be
\tau\propto\frac{\Gamma d^2}{G}\sim\frac{\eta d^2}{G\xi^2}\label{tau}
\ee
for the radial motion with axial profile $u_r\propto\sin(\pi z/d)$. The simplified linear equation of motion is 
\be
\Gamma{\dot u_r}\simeq -(\pi^2 G/d^2)u_r.\label{urcomp2}
\ee

What is still missing from this analysis is the additional force in \eqref{hoop} acting on the network due to hoop stresses. Being fundamentally nonlinear, this is not captured by \eqref{2fluidElastic}, so we add this to the results of the (linear) two-fluid model to obtain the following phenomenological equation of motion:
\be
\nabla_rP=\Gamma\dot u_r=-K\frac{u_r}{r^2}-\frac{1}{r}\tilde AG\gamma^2.\label{2Fluid}
\ee
Here, $K\simeq\pi^2G/\tan\left(\alpha\right)^2$ and we have used the fact that $d=\tan\left(\alpha \right)r$.
\sn{Importantly, this is a linear equation of motion for $u_r$, in spite of the $\sim\gamma^2$ driving force. The latter is controlled by the rheometer and acts as in inhomogeneous term in an otherwise linear differential equation for the unknown $u_r(t)$.}

\subsection*{Incompressible limit}
First, let's consider the case of an incompressible medium, as one has for the limit $\Gamma\rightarrow\infty$, where $u\rightarrow0$ and
\be
\nabla_rP=-\frac{1}{r}\tilde\sigma.
\ee
This equation can be integrated to give
\be
P(R)-P(r)=-\tilde\sigma\log(R/r),
\ee
where $P(R)$ is the pressure at the sample boundary, i.e., atmospheric pressure $P_0$. 
The excess pressure,
\be
\Delta P=P(r)-P_0
\ee
can be integrated to give a positive (upward) contribution to the thrust $F$ 
\be
2\pi\tilde\sigma\int_0^R r\log(R/r)dr=\frac{\pi R^2}{2}\tilde\sigma.
\ee
Adding this to the direct contribution
\be
-\pi R^2\sigma_{zz}
\ee
from 
$\sigma_{zz}$, we find that
\be
\frac{2F}{\pi R^2}=N_1=\sigma_{xx}-\sigma_{zz},\label{N1-incomp}
\ee
implying that, 
\be
\tilde\sigma=\sigma_{xx}+\sigma_{zz}\simeq \tilde AG\gamma^2,
\ee 
where $\tilde A=\left(A_x+A_z\right)$, 
\be
\sigma_{xx}\equiv A_xG\gamma^2\quad\mbox{and}\quad\sigma_{zz}\equiv A_zG\gamma^2.
\ee
We expect $A_{z}\sim 1/\gamma_0$, based on the prior low-frequency model \cite{Janmey2007,Kang2009}.
Since $\sigma_{xx}$ usually is of order but larger than $\sigma_{zz}$, we expect both
$A_{x,z}\sim 1/\gamma_0$, which is typically of order 10 for biopolymer networks. Moreover, as defined both stress components are strictly positive (tensile) and we expect that $A_x>A_z$, for $N_1>0$ in the incompressible limit.

\subsection*{Compressible limit}
In the limit of long times $t\gg\tau$ and low frequencies $\omega\tau\ll 1$ in Eq.\ (\ref{2Fluid}), the pressure vanishes, and the apparent $N_1$ measured is that of
Refs.\ \cite{Janmey2007,Kang2009}
\be
\frac{2F}{\pi R^2}=N_1^{\mbox{\scriptsize app}}=-2\sigma_{zz}=-2A_zG\gamma^2.\label{N1-comp}
\ee
This describes the long-time value in Fig.\ 2c of the main text, to which the normal stress relaxes. 
For intermediate times/frequencies, we solve Eq.\ (\ref{2Fluid}), with $\gamma(t)=\gamma_0\sin(\omega t)$, to find both steady-state (ss) and transient (tr) solutions, where the general $u_r(t)=u^{\mbox{\scriptsize (ss)}}(t)+u^{\mbox{\scriptsize (tr)}}(t)$. 
We find 
\be
u^{\mbox{\scriptsize (ss)}}(t)
=-\frac{\tilde AG_0\gamma_0^2 r \left(-K^2 \cos (2 t \omega )+K^2-2 \Gamma  K r^2 \omega  \sin
	(2 t \omega )+4 \Gamma ^2 r^4 \omega ^2\right)}{2 \left(K^3+4 \Gamma ^2
	K r^4 \omega ^2\right)}
\ee
and
\be
u^{\mbox{\scriptsize (tr)}}(t)=\frac{\tilde AG_0\gamma_0^2 r}{2 K \left(\frac{K^2}{4 \Gamma ^2 r^4 \omega ^2}+1\right)}e^{-\frac{K}{\Gamma  r^2}t}.
\ee
The transient is found by choosing a homogeneous solution of \eqref{2Fluid} to give $u_r(t)=u^{\mbox{\scriptsize (ss)}}(t)+u^{\mbox{\scriptsize (tr)}}(t)=0$ at $t=0$. 
The transient is most relevant to the case where its characteristic relaxation time $\tau$ is large compared with the period of oscillation $\sim1/\omega$. Thus, we approximate
\be
u^{\mbox{\scriptsize (tr)}}(t)\simeq\frac{\tilde AG_0\gamma_0^2 r}{2 K}e^{-\frac{K}{\Gamma  r^2}t},
\ee
from which we find
\be
\nabla_rP=\Gamma\dot u_r\simeq-\frac{\tilde AG_0\gamma_0^2}{2 r}e^{-\frac{K}{\Gamma  r^2}t}.\label{last}
\ee
As noted before, there is no single relaxation time in this system, since $\tau\sim\eta d^2/(G\xi^2)$ depends on the gap $d$. This can be seen in \eqref{last}, where the relaxation is $r$-dependent. This can be integrated to find the transient contribution to $N_1^{\mbox{\scriptsize app}}$ given by
\be
N_1^{\mbox{\scriptsize transient}}\simeq\frac{1}{2} \tilde AG_0\gamma_0^2 \left(\frac{ \text{Ei}\left(-t/\tau\right)t}{\tau}+e^{-t/\tau}\right)
\simeq\frac{1}{2} \tilde AG_0\gamma_0^2\exp\left[{-1.91\left(\frac{t}{\tau}\right)^{0.78}}\right],\label{taufit}
\ee
where the latter approximation is valid to within less than 2\% until the transient has decayed to less than 2\% of its initial value. 

For the steady-state,  
\be
N_1^{\mbox{\scriptsize app}}=-2A_zG\gamma^2+\tilde AG\gamma_0^2 \left({\mathcal A}\cos (2 \omega t)+{\mathcal B}\sin (2\omega t)\right),\label{N1app}
\ee
where
\be
{\mathcal A}=-\frac{1}{8 \omega\tau  }
\Bigg[2 \tan ^{-1}\left(1+2 \sqrt{\omega\tau}\right)
+2 \tan ^{-1}\left(1-2 \sqrt{\omega\tau}\right)
-\pi +4 \omega\tau\Bigg]\label{calA}
\ee
and
\be
{\mathcal B}=\frac{1}{8\omega\tau }\log\left(1+4\omega^2\tau^2\right).\label{calB}
\ee
For an incompressible system, \eqref{N1-incomp} is recovered for $u_r(t)=u^{\mbox{\scriptsize (ss)}}(t)+u^{\mbox{\scriptsize (tr)}}(t)$ as $\tau\rightarrow\infty$.
In the limit of low frequency, this reduces to the fully compressible limit of \eqref{N1-comp}. 

\subsection*{Application of the model to the experimental results}
\begin{figure}[htbp]
	\centering
	\includegraphics[width=0.4\textwidth]{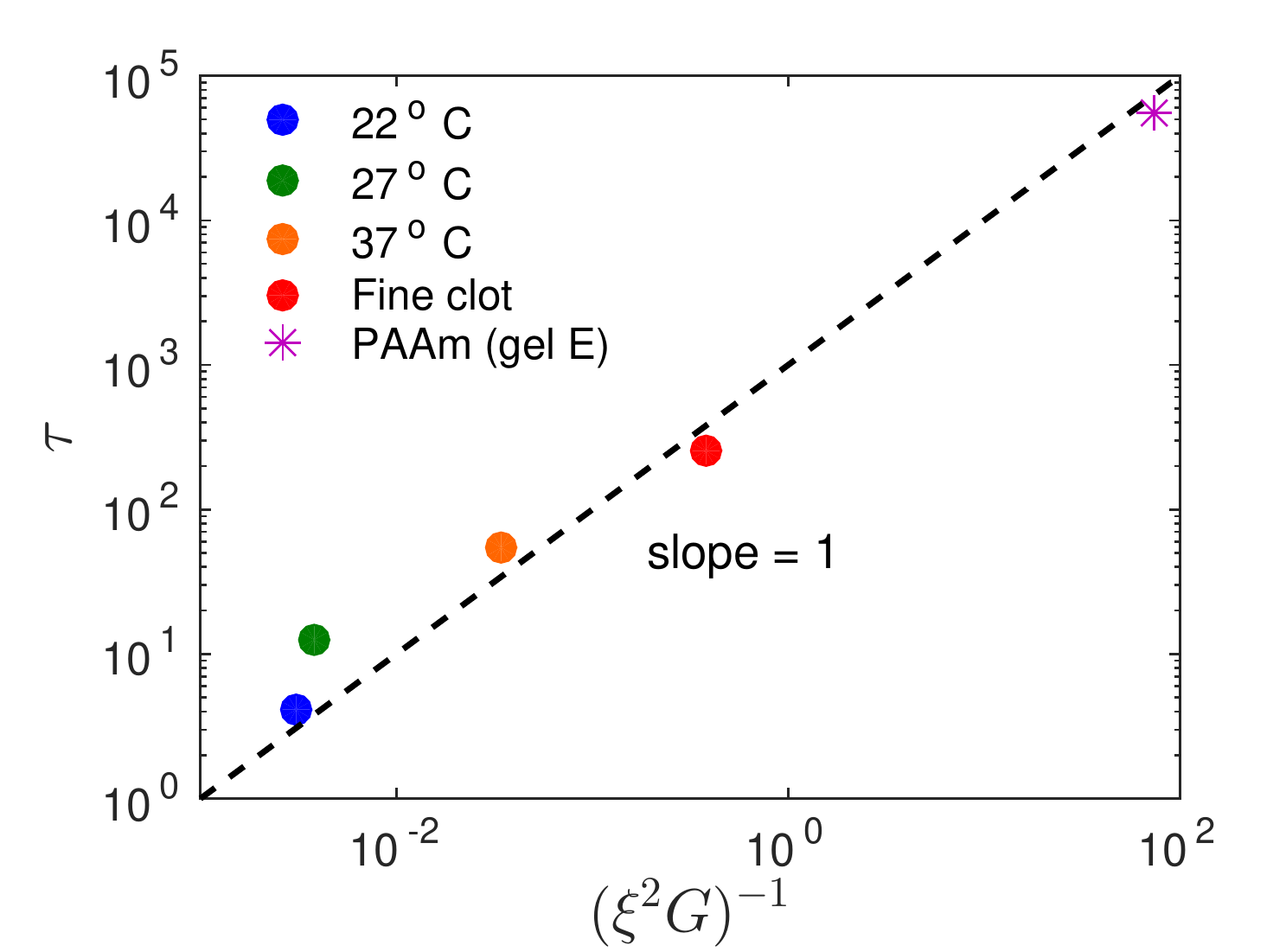}
	\includegraphics[width=0.4\textwidth]{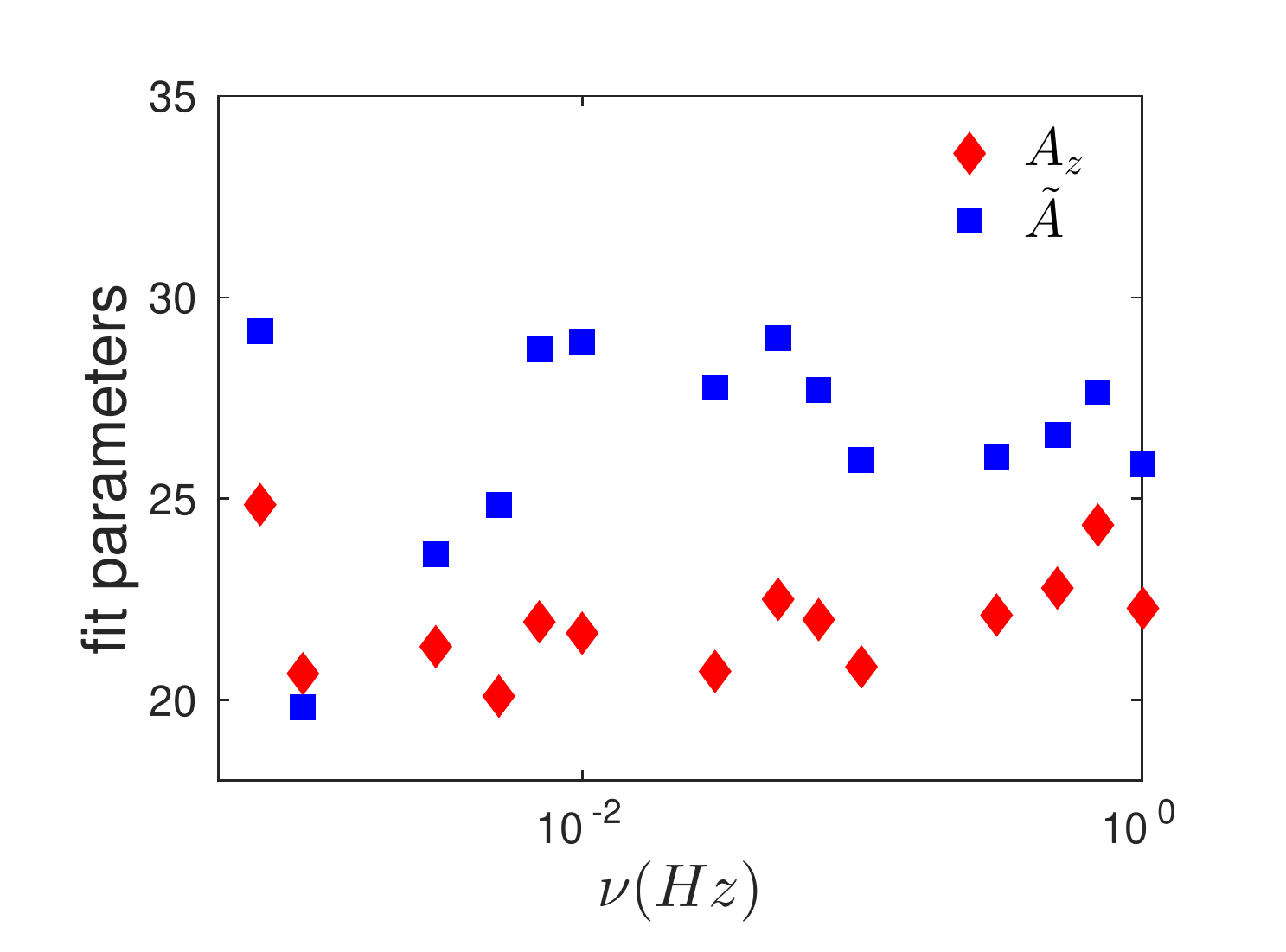}

	\caption{ \sn{a) Relaxation times for fibrin and PAAm gels are obtained from the fits in Fig.\ 2 of main text. These relaxation times are plotted vs $1/(G\xi^2)$ for comparison with predicted relaxation time dependence on $\xi$ and $G$. For fibrin gels see SI pore size analysis below. For PAAm, we use $\xi=10$nm, which is consistent both with prior literature values \cite{tombs1965} and with the estimated $\xi\sim\sqrt{\eta d^2 / (G \tau)}$ using our model, together with the fit $\tau=15.5$h, $G=141$Pa, $\eta=10^{-3}$Pa and $d=R\tan(\alpha)=25 \tan(2^\circ)$mm $\simeq 1$mm.} Fit parameters ($A_z$ and $\tilde{A}$) for the fibrin sample polymerized at $27^{\circ}$C versus frequency.}
	\label{fig:fitparam}
\end{figure}

The experiments are done on fibrin samples with different mesh sizes, obtained by varying the temperature, ionic conditions and pH. To explain the experimental results using the two-fluid model, we fit the steady state normal stress data to Eq. (\ref{N1app}). The free parameters in this equation are $A_z$ and $\tilde{A}$. In this formula $\omega$ is the frequency of oscillatory shear stress or strain. The other parameters $G$, $\gamma_0$ and $\tau$ can all be directly obtained experimentally. The amplitude of the shear strain, $\gamma_0$ is evaluated by fitting a sinusoidal function to the recorded shear strain data. The linear shear modulus $G$ is obtained by fitting a linear stress-strain relation to the stress-strain curves. In all cases, the relaxation time $\tau$ is obtained by fitting the normal stress relaxation curves versus time to Eq. (\ref{taufit}), \sn{as shown in Fig.\ 2. The results are plotted versus $1/(G\xi^2)$, to test the predicted scaling $\xi$ and $G$. The mesh size for fibrin samples has been measured (see SI pore size analysis), while for PAAm we use a mesh size of 10 nm, consistent with prior literature \cite{tombs1965} and the predicted $\tau\sim\eta d^2/G\xi^2$ using $G=141$Pa, $\eta=10^{-3}$Pa and $d\simeq 1$mm.} For fibrin gels polymerized at $27^\circ$ C, we find an average relaxation time $\tau=12.5$s \sn{and $G=3150$ Pa}. We apply frequencies in the range 0.001 Hz to 1 Hz. In Fig. \ref{fig:fitparam}, the fit parameters from the two-fluid model are plotted versus frequency. In Fig. 3 of the main text, we have chosen three frequencies (0.001, 0.01 and 1 Hz) out of this frequency range, to show how the normal stress response changes from extensile to contractile behavior with changing frequency.

\newpage

\section*{Transient regime during oscillatory shear measurements}

\begin{figure}[h] 
\centering 
\noindent\makebox[\textwidth]{\includegraphics[width=8cm]{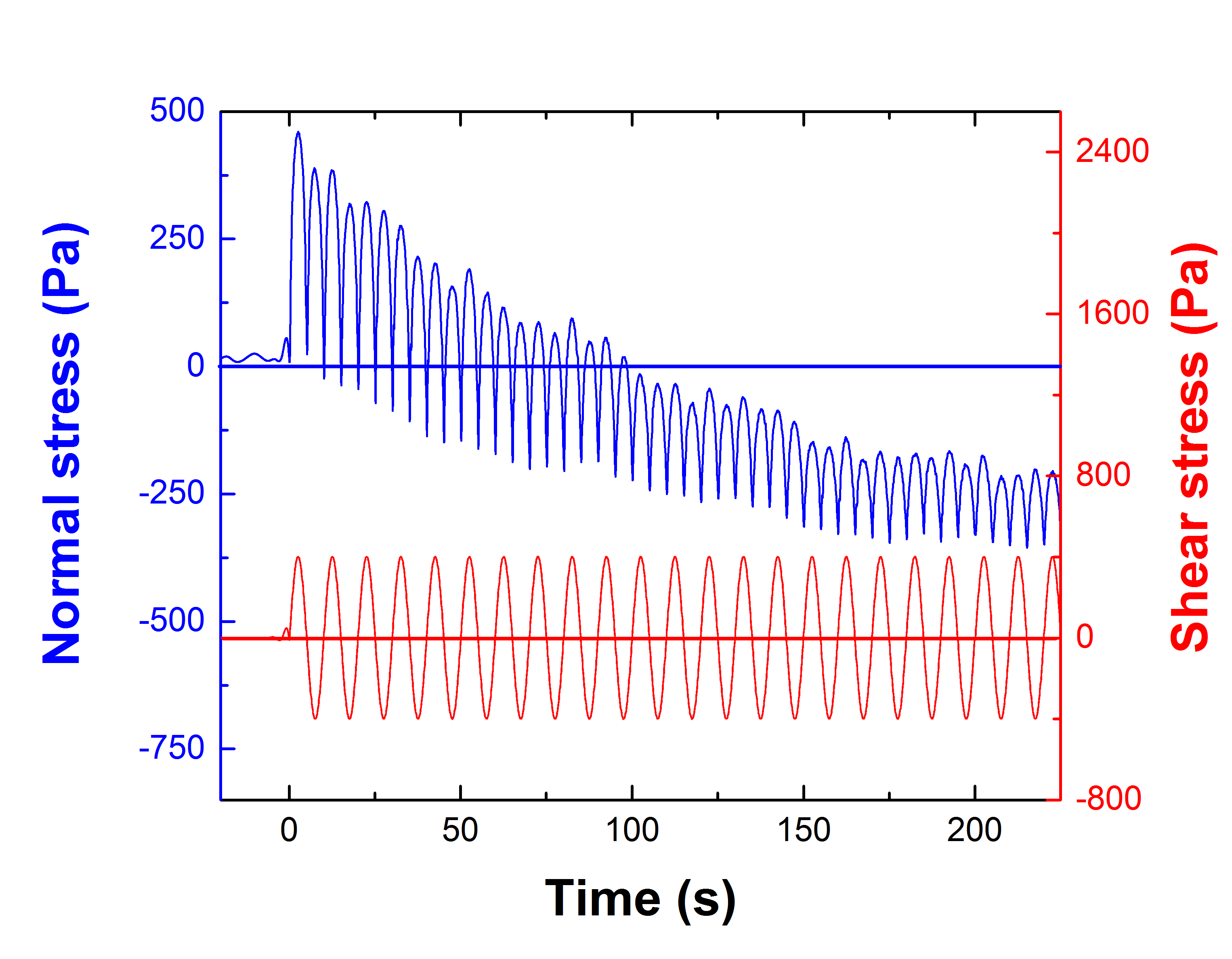}} 
\caption{Transient from a positive to a negative normal force for a fibrin gel, polymerized under fine clotting conditions, under oscillatory shearing ($f=0.1Hz, \sigma=400Pa$)}
\label{Fn_Transient}
\end{figure}

To measure the transition time in a fibrin gel polymerized under fine clot conditions (pore size $0.08\ \mu$m), we applied an oscillatory shear stress and measured the normal force as a function of time. Fig. \ref{Fn_Transient} shows the transient regime, where the normal stress relaxes from a positive to a negative value over a characteristic time scale $\tau$. This relaxation is similar ($\tau = 100$ s) to what is observed for experiments where a constant shear stress is applied (as shown in the main text in Figure 2).

\newpage

\section*{Influence of rheometer shear cell geometry on the normal stress signal}

In this section, we show normal stress measurements obtained by shearing fibrin gels polymerized at 27 \textdegree C with different geometries. The dimensional analysis explained in the section 'Two-fluid model and relaxation of hoop stress' predicts that the characteristic time constant of the gel should scale as $\tau\sim \frac{\eta d^2}{G\xi^2}$ (Eq.\ (\ref{tau})), where $d$ is a characteristic length scale of the problem. By changing the measurement geometries, we thus expect the normal force signal at a given frequency to change.

\begin{figure}[h] 
	\centering 
	\noindent\makebox[\textwidth]{\includegraphics[width=15cm]{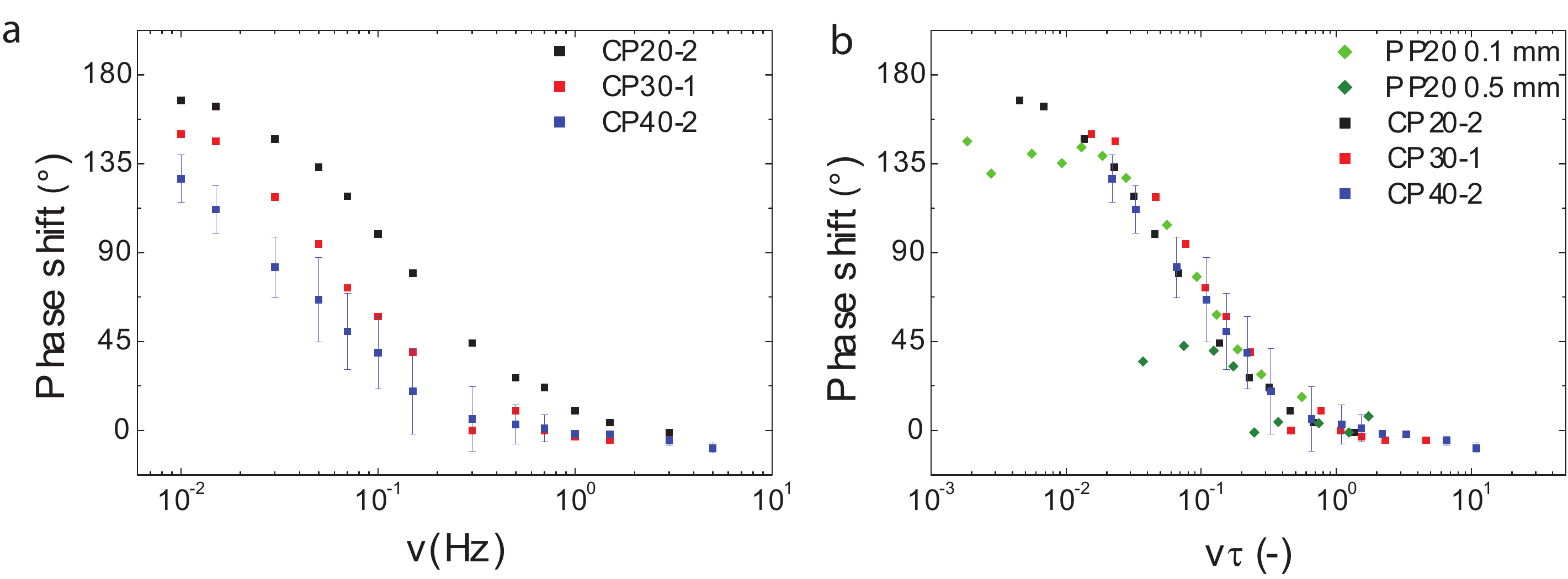}} 
	\caption{\textbf{a}: Apparent phase shift between the normal stress and the shear strain for fibrin gels polymerized at 27\textdegree C as a function of the shearing frequency, for various shearing geometries. \textbf{b}: The frequency axis is rescaled by the characteristic time $\tau\sim\eta d^2/(G\xi^2)$. Error bars on the CP40-2 data set represent the standard deviation between 6 measurements.}
	\label{SIgeometries}
\end{figure}

We measure for different frequencies the phase shift $\phi$ between the normal force signal, $N(t)=A\cos(2.(2\pi\nu) t + \phi)$, and the squared shear strain, $\gamma(t)^{2}=(\gamma_{0}\cos((2\pi\nu).t))^{2}=\frac{\gamma_{0}^{2}}{2}(1+\cos(2.(2\pi\nu).t)$. We choose to compare $N(t)$ with $\gamma(t)^{2}$ because both quantities have the same frequency ($2\nu$), and because of the analogy with the Mooney-Rivlin model prediction, $N=G\gamma^{2}$. The phase shift as a function of frequency measured for several plate-plate (PP) geometries and cone-plate (CP) geometries is shown in Fig. \ref{SIgeometries}. We observe a marked dependence on gap size and cone angle.

As shown in Fig.~\ref{SIgeometries}b, we can scale out these differences by rescaling the frequency axis by the characteristic relaxation time $\tau\sim\frac{\eta d^2}{(G\xi^2)}$, where $\eta$ is the viscosity of the interstitial fluid ($10^{-3}$ Pa s), $G$ is the storage modulus of the gel (which varies slightly with each experiment), $d$ is the gap of the rheometer (chosen at the edges for the cone-plate geometries) and $\xi$ is the mesh size of the gel (0.29 $\mu$m at 27\textdegree C). We observe a collapse of all the curves (Fig.~\ref{SIgeometries}b), showing that the gap size is the relevant length scale that governs the time-dependence of the normal force, in accordance with predictions from the dimensional analysis.

\newpage

\section*{Pore size analysis}

The pore size of the fibrin gels was determined using turbidimetry, and verified by confocal microscopy. Turbidimetry data was analyzed based on a model for light scattering from isotropic networks of rigid rods \cite{Piechocka2016,Yeromonahos2010} in order to obtain the average mass-length ratio of the fibrin fibers. The fiber density in terms of the total length per volume, $\rho$, was calculated from the mass-length ratio and the fibrinogen mass concentration; we then obtained the mesh size $\xi\sim\sqrt{1/\rho}$.

\begin{figure}[h] 
	\centering 
	\noindent\makebox[\textwidth]{\includegraphics[width=10cm]{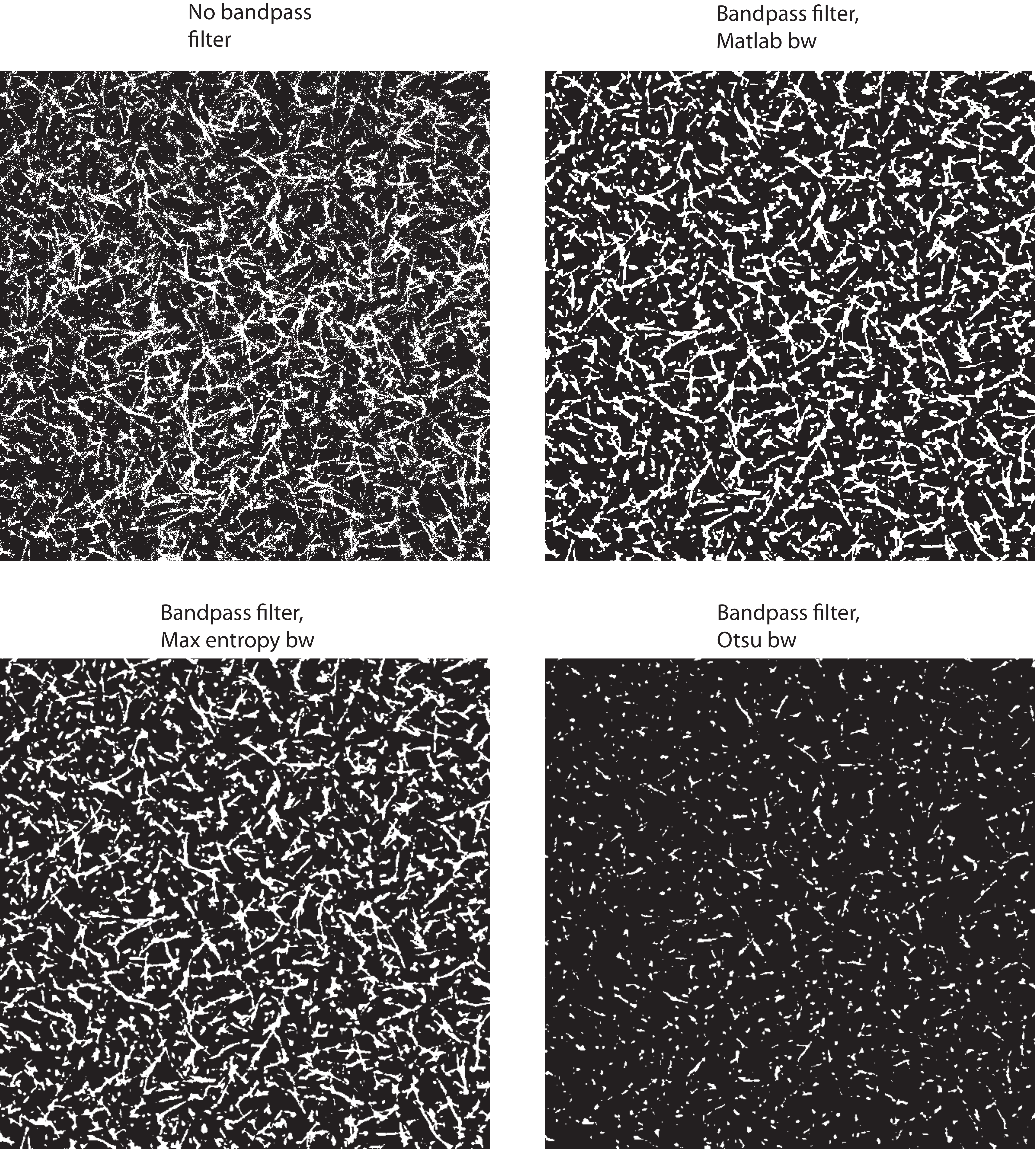}} 
	\caption{Various thresholding methods (Matlab's default thresholding function, Kapur's or Maximum Entropy method \cite{Kapur1985}, Otsu's method \cite{Otsu1979}) applied on the same confocal plane, without (top left) or with a FFT bandpass filter. Images are 40 by 40 $\mu$m.}
	\label{Thresholding}
\end{figure}

\begin{figure}[h] 
	\centering 
	\noindent\makebox[\textwidth]{\includegraphics[width=12cm]{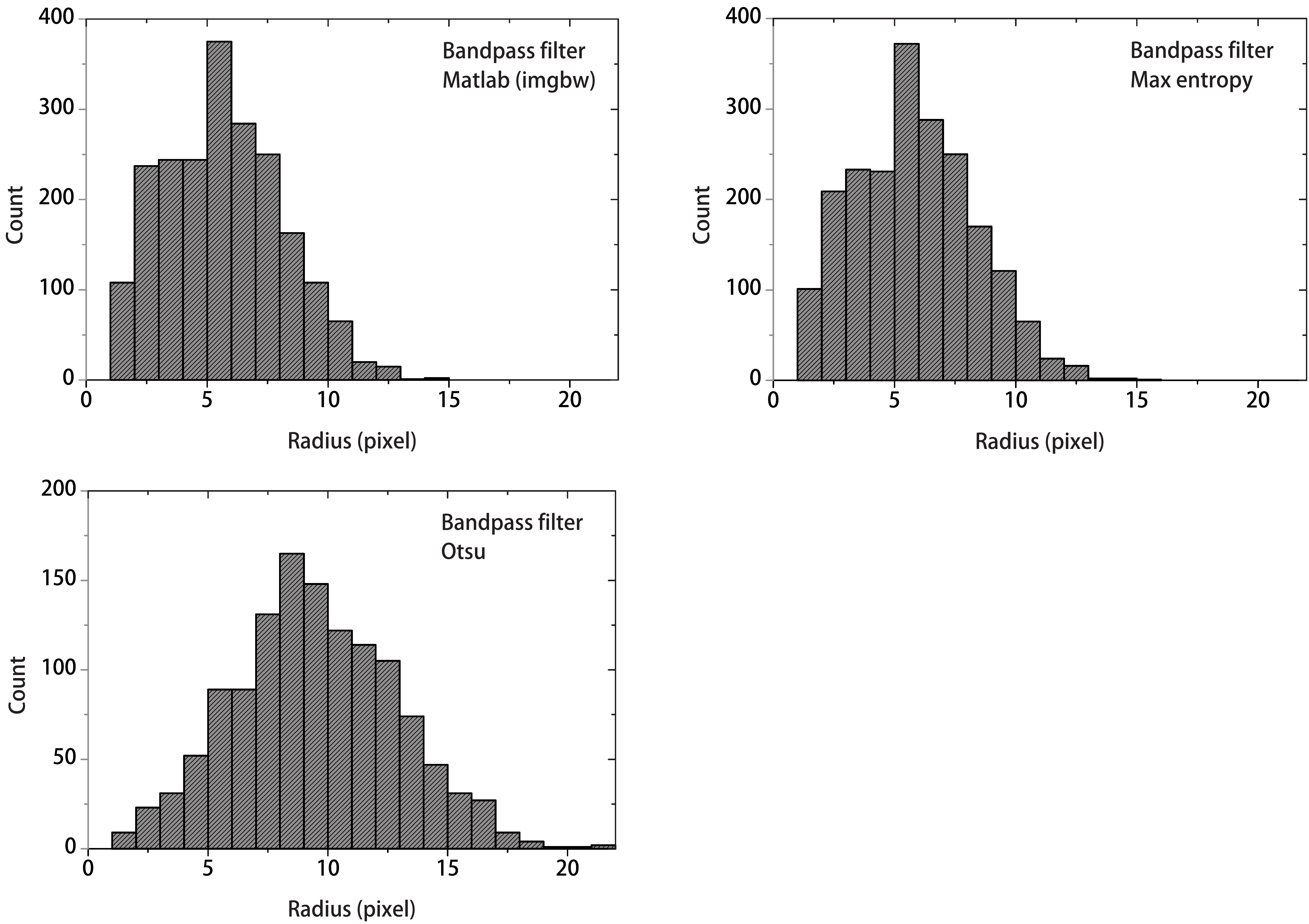}} 
	\caption{Pore size distribution resulting from bubble analysis as described in \cite{Munster2013}, applied on images that are thresholded using various methods.}
	\label{Pore_size}
\end{figure}

An independent way to obtain the gel's pore size is through analysis of confocal images \cite{Munster2013}. Analysis was done on 2 independently polymerized samples. Planes were analyzed separately. A FFT bandpass filter was applied to each confocal plane, filtering out features smaller than 2 pixels and larger than 30 pixels. As fibers were typically 5 pixels in diameter, this filtering step preserves their structure. Various thresholding techniques were applied: the built-in Matlab thresholding function (imgbw), Kapur's thresholding method (also known as the Maximum Entropy method) \cite{Kapur1985} and Otsu's thresholding method \cite{Otsu1979}. An overview of images with the FFT filter and thresholding applied is shown in Figure \ref{Thresholding}. The pore sizes obtained using analysis of turbidity spectra and image analysis are shown in Table \ref{Pore_size_table}. Although absolute values differ, the ratios between the 22\textdegree C and the 27\textdegree C show a consistent picture where the pore size of the fibrin network decreases upon an increase in the polymerization temperature by a factor of approximately 1.2. A histogram of the resulting pore size distributions is shown in Figure \ref{Pore_size}.



\begin{table}[]
	\centering
	\begin{tabular}{|l|l|l|l|}
		\hline
		& 22\textdegree C, pore size ($\mu$m) & 27\textdegree C, pore size ($\mu$m) & Ratio \\ \hline
		BP filter, Matlab (imgbw)     & 0.66 (0.00)   & 0.54 (0.03)   & 1.21      \\ \hline
		BP filter, Max. entropy       & 0.68 (0.01)   & 0.60 (0.03)   & 1.12      \\ \hline
		BP filter, Otsu               & 1.19 (0.10)   & 0.92 (0.12)   & 1.29      \\ \hline
		Photospectrometry             & 0.36 (0.01)   & 0.29 (0.01)   & 1.24      \\ \hline
	\end{tabular}
	\caption{Pore sizes in $\mu m$ of 8 mg/ml fibrin gels polymerized at 22\textdegree C and 27\textdegree C, obtained using bubble analysis of confocal images with various thresholding methods, and turbidimetry. The number between brackets is the standard deviation between measurements.}
	\label{Pore_size_table}
\end{table}

\newpage

\bibliographystyle{ieeetr}
\bibliography{References}

\end{document}